\documentclass[prl,twocolumn,showpacs,superscriptaddress]{revtex4}
\usepackage{amsmath,amssymb,graphicx,epstopdf,color,bm,soul,upgreek}

\begin{document}

\title{Spatial Coherence Properties of One-Dimensional Exciton-Polariton-Condensates}

\author{J. Fischer}
\affiliation{Technische Physik, Wilhelm-Conrad-R\"ontgen-Research Center for Complex Material Systems, Universit\"at W\"urzburg, Am Hubland, D-97074 W\"urzburg, Germany}

\author{I. G. Savenko}
\affiliation{QCD Labs, COMP Centre of Excellence, Department of Applied Physics and Olli V. Lounasmaa Laboratory, Aalto University, P.O. Box 13500, FI-00076 Aalto, Finland}
\affiliation{National Research University of Information Technologies, Mechanics and Optics (ITMO), St. Petersburg 197101, Russia}

\author{M. D. Fraser}
\affiliation{Quantum Optics Research Group, RIKEN Center for Emergent Matter Science, 2-1 Hirosawa, Wako-shi, Saitama 351-0198, Japan}

\author{S. Holzinger}
\affiliation{Technische Physik, Wilhelm-Conrad-R\"ontgen-Research Center for Complex Material Systems, Universit\"at W\"urzburg, Am Hubland, D-97074 W\"urzburg, Germany}

\author{S. Brodbeck}
\affiliation{Technische Physik, Wilhelm-Conrad-R\"ontgen-Research Center for Complex Material Systems, Universit\"at W\"urzburg, Am Hubland, D-97074 W\"urzburg, Germany}

\author{M. Kamp}
\affiliation{Technische Physik, Wilhelm-Conrad-R\"ontgen-Research Center for Complex Material Systems, Universit\"at W\"urzburg, Am Hubland, D-97074 W\"urzburg, Germany}

\author{I. A. Shelykh}
\affiliation{Science Institute, University of Iceland, Dunhagi 3, IS-107, Reykjavik, Iceland}
\affiliation{Division of Physics and Applied Physics, Nanyang Technological University 637371, Singapore}

\author{C. Schneider}
\affiliation{Technische Physik, Wilhelm-Conrad-R\"ontgen-Research Center for Complex Material Systems, Universit\"at W\"urzburg, Am Hubland, D-97074 W\"urzburg, Germany}

\author{S. H\"ofling}
\affiliation{Technische Physik, Wilhelm-Conrad-R\"ontgen-Research Center for Complex Material Systems, Universit\"at W\"urzburg, Am Hubland, D-97074 W\"urzburg, Germany}
\affiliation{SUPA, School of Physics and Astronomy, University of St Andrews, St Andrews KY16 9SS, United Kingdom}

\begin{abstract}
In this work, we combine a systematic experimental investigation of the power- and temperature-dependent evolution of the spatial coherence function, $g^{\left(1\right)}\left(\boldsymbol{r}\right)$, in a one-dimensional exciton-polariton channel with a modern microscopic numerical theory based on a stochastic master equation approach. The spatial coherence function $g^{\left(1\right)}\left(\boldsymbol{r}\right)$ is extracted via high-precision Michelson interferometry, which allows us to demonstrate that in the regime of non-resonant excitation, the dependence $g^{\left(1\right)}\left(\boldsymbol{r}\right)$ reaches a saturation value with a plateau, determined by the intensity of pump and effective temperature of the crystal lattice. The theory, which was extended to allow for treating incoherent excitation in a stochastic frame, matches the experimental data with good qualitative and quantitative agreement.  This allows us to verify the prediction that the decay of the off-diagonal long range order can be almost fully suppressed in one-dimensional condensate systems.   
\end{abstract}

\maketitle

{\it Introduction.---} Exciton-polaritons evolve in semiconductor microcavities as the result of strong coupling of optical and matter modes~\cite{kavbamalas}. At sufficiently low concentrations (up to $\sim 10^{11}$ cm$^{-2}$) they obey bosonic statistics, and owing to their small effective mass, which is about $10^{5}$ times smaller than the free electron mass, manifest quantum coherent properties at surprisingly high temperatures. Indeed, polariton condensation is observed at temperatures of tens of Kelvin in GaAs and CdTe-based structures \cite{KasprzakNature,BaliliScience} and up to room temperatures in wide-bandgap materials \cite{Christopolous2007,Deng2012,Mahrt2013}.
Having lot of similarities with conventional Bose-Einstein Condensation (BEC) \cite{kavbamalas}, polariton condensation reveals some important peculiarities. Differently from cold atoms, polaritons have finite lifetime and in order to reach an equilibrium state their radiative decay should be compensated by a constant pumping of the system, which can be implemented either optically or electrically \cite{Schneider2013}. Moreover, the planar microcavity exciton-polariton system is inherently two-dimensional (2D), and in accordance with the Hohenberg-Mermin-Wagner theorem the transition to BEC in a uniform system is only possible at zero temperature \cite{hohenberg} for 2D as well as 1D geometries. Restriction of the system to a finite size, however, inhibits excitation of density and phase fluctuations permitting the formation of a condensate or quasi-condensate phase with a macroscopic coherence length \cite{petrov2000bose, petrov2000regimes}. 

%Therefore, one can expect that polaritons can only form spots of quasi-condensate with a finite coherence length (usually of the order of the optical pumping spot size~\cite{KasprzakNature,Richard2005}), and at macroscopic lengths the system enters the Berezinskii-Kosterlitz-Thouless phase \cite{Nitsche-arxiv14}.

There is a long standing discussion in the literature addresing the question: what should be the experimentally verifiable criterion of the polariton condensation? It is now commonly accepted, that the onset of the Off-Diagonal Long Range Order (ODLRO), determined by the first-order spatial coherence function $g^{\left(1\right)}\left(\textbf{r}\right)$ can be considered as a smoking gun criterion, putting the latter function in the most intense focus of theoretical \cite{Sarchi2007,PolcohDoan,SavenkoJETP2012,WoutersArxiv} and experimental \cite{KasprzakNature,SpatCohPol,Manni2011} research. In particular, polariton condensation in one-dimensional channels has drawn special attention. It was demonstrated that polaritons trapped in a channel reveal the appearance of ODLRO in the condensation regime \cite{Manni2011}. Moreover, the droplets of BECs in such systems can propagate over macroscopic distances, preserving their coherence properties and allowing for the efficient manipulation~\cite{RefTosi2012}. Recently, an experimental observation of room-temperature polariton condensation in a one-dimensional ZnO channel has been reported~\cite{Trichet2011}. These technological achievements open a route to the creation of polariton-based logic elements and optical integrated circuits working at relatively high temperatures~\cite{LiewIntegrated}. Among the phases expected for a 1D polariton gas are the Tonks gas \cite{paredes2004} and the condensate/quasi-condensate phases \cite{petrov2000regimes}, the former appearing when interactions come to dominate the energy scale and the condensate phases resulting for a sufficiently weakly-interacting and finite-sized system. The quasi-condensate which still exhibits some phase fluctuations is characterised by an exponentially decaying long-range order, but with enhanced coherence length.  Such a phase is typical of the weakly-interacting dilute-gas atomic condensate.  The low energy of phase fluctuations in this system prevents the formation of a true condensate phase with a plateauing long-range order, except at the lowest temperatures \cite{petrov2000regimes}. 
The decay of spatial coherence in a low-dimensional non-equilibrium (polariton) condensate has been studied through a mean-field treatment with a stochastic noise term \cite{chiocchetta2013non} with the result that, provided the gas is interacting, similar behaviour to that of equilibrium BEC is expected in 1D, 2D and 3D systems.
The results that we discuss in this letter however distinctly place our 1D polariton condensate into the true BEC-like phase where both phase and density fluctuations are suppressed to the level that ODLRO can be observed with a BEC-like constant plateau in the spatial coherence at the largest distances. \\
On the theoretical side, for the description of the polariton dynamics in one dimension an approach based on the Lindblad master equation technique has been developed \cite{SavenkoPRB2011}. Differently from the approaches based on either Gross-Pitaevskii-type equations \cite{Carusotto2004,Shelykh2006} or semi-classical Boltzmann equations \cite{Porras2002,Haug2005,Cao2008}, it allows to account for real space dynamics of the polariton droplets, processes of decoherence and energy relaxation provided by the interaction of polaritons with the thermal bath of acoustic phonons. Moreover, the calculation of two-point correlators necessary for the analysis of the transition between thermal and OLDRO phases became possible \cite{SavenkoJETP2012}.

Here, we demonstrate that once the intensity of pump reaches a thres\-hold value, the system undergoes the transition from the thermal to condensate phase accompanied by the onset of ODLRO. Theoretically, this transition is modeled using the original microscopic approach based on a stochastic density matrix for\-ma\-lism. We investigate the influence of temperature and pump intensity modulation on the coherence properties of the system and reveal good agreement between the theory and experiment.\\

%---------------------------
%---------------------------
%---------------------------

\begin{figure}[tbp]
\centering
\includegraphics[width=\linewidth]{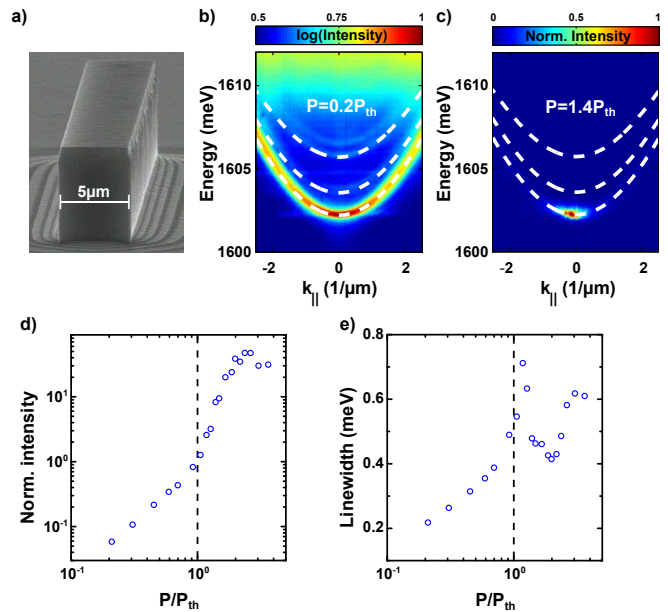}
\caption{a) Scanning electron microscope image of a one-dimensional polariton channel. b) Power dependent Fourier-space spectrum of the microwire emission below,  and c) above the regime of polariton condensation. d) Input-output characteristics of the ground state emission from the wire. e) Polariton linewidth as a function of the pump power. }
\label{fig:Fig01}
\end{figure}

{\it Experiment.---} The polariton channels (Fig. \ref{fig:Fig01}a)) with a length of 200 µm (width 5 µm) were etched into a high-Q AlGaAs based $\lambda/2$ microcavity with twelve GaAs quantum wells (QWs) and a Q-factor exceeding 10000 (see supplementary information \cite{Supplement}).  

First, we investigate the power dependent emission features of the microwire cavity via momentum resolved spectroscopy. The microwire is mounted in a helium flow cryostat and it is oriented pa\-ral\-lel to the entrance slit of the spectrometer. Polaritons are injected with a non-resonant CW-laser which is tuned to the energy of the first reflection minimum of the Bragg reflector. The Gaussian-shaped pumping spot has a diameter of 5\,$\upmu$m, and the laser beam is chopped with a duty cycle of 0.1 to reduce power-induced heating effects. As shown in Fig. \ref{fig:Fig01}b), we observe a set of parabolic dispersions which can be attributed to the ground state of the microwire as well as higher order lateral modes. Each dispersion is split into a mode doublet \cite{Kuther1998}. The energy-momentum detuning between the lowest photon mode $E_{C1}\left(k_{\parallel}=0\right)$ and the exciton energy $E_{X}\left(k_{\parallel}=0\right)$ of the ground state amounts to $\delta=E_{C1}\left(k_{\parallel}=0\right)-E_{X}\left(k_{\parallel}=0\right)=-13$\,meV, which is less than the Rabi splitting $E_{RS} =14.5$\,meV. With the increase of the pump power, polaritons condense into the ground state which is manifested by its massive occupation [Fig. \ref{fig:Fig01}c)]. The corresponding input-output characteristics of the power dependent study is shown in Fig. \ref{fig:Fig01}d). It features a distinct threshold which is typical for the onset of stimulated scattering leading to the growth of a condensate. Above $P=2.5P_{th}$ the intensity decreases again indicating excitation power-induced heating of the sample. At threshold, the linewidth of the polariton emission (shown in Fig. \ref{fig:Fig01}e) significantly reduces. This behaviour is commonly assigned to an increased temporal coherence of the emission in the regime of polariton lasing \cite{Tempel2012}. Further, it increases again with the increase of the particle densities which can be attributed to dephasing resulted from the polariton-polariton interactions.

\begin{figure}[tbp]
\centering
\includegraphics[width=\linewidth]{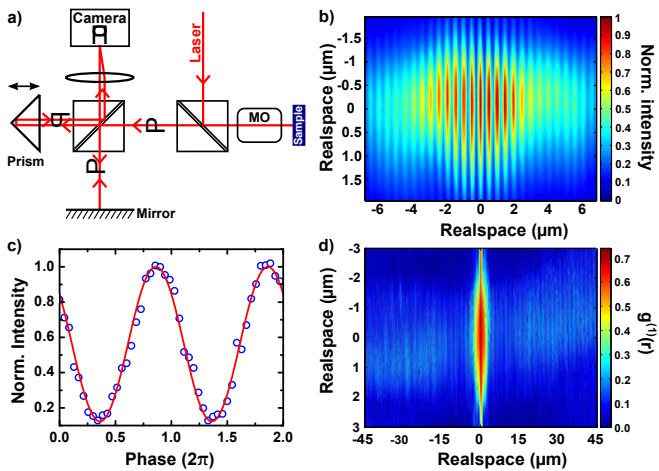}
\caption{a) Schematic sketch of the Michelson-interferometer setup for the determination of the spatial correlation function. b) Typical interference pattern at an excitation power of $P=1.6P_{th}$. c) By variation of the distance between the mirror and prism the intensity of on pixel undergoes a sine-function (blue circles). The red line shows the fitted sine from which the $g^{\left(1\right)}$-value and the phase can be calculated. d) $g^{\left(1\right)}$-map received by fitting all pixel of the camera image with a sine-function above condensation threshold at $P=1.6P_{th}$.}
\label{fig:Fig02}
\end{figure}

In order to investigate the spatial coherence properties of the polariton channels, we measure its spatial correlation function, $g^{\left(1\right)}(\textbf{r}',\textbf{r}'')$. We use a Michelson interferometer with a variable path length which overlaps the real space image of the polariton emission with its mirror image generated by a right angle prism. A schematic drawing of the optical setup is shown in Fig. \ref{fig:Fig02}a) (it is similar to ref. \cite{Roumpos-PNAS12}). The overlapped real space images from the polariton wire are combined onto a CCD camera with high spatial resolution. The light from the excitation laser is filtered out with a long pass filter in the optical beam path. Fig. \ref{fig:Fig02}b) depicts the resulting interference pattern on the camera. By moving one arm of the interferometer, we can extract the interferograms. From the visibility of the interference fringes [see Fig. \ref{fig:Fig02}c)], we can reconstruct the full spatial correlation function $g^{\left(1\right)}(r,-r)$ which is plotted in the colour map in \ref{fig:Fig02}d). It is interesting that, in contrast to the experimentally reported algebraic decay in 2D polariton systems \cite{Roumpos-PNAS12,SpatCohPol}, the spatial coherence function of a one-dimensional condensate is indicated to reach a constant plateau at large distances \cite{Bloch2011,Trichet2013,Manni2011}. However, in these reports the clarity of this behavior is either obscured by large experimental errors resulting from the applied double slit technique \cite{Bloch2011} or by strong oscillations in the correlation function related to inhomogeneities of the 1D trap \cite{Manni2011} and a comparably short long range order in the range of 10\,$\upmu$m \cite{Trichet2013}. 
For the purposes of assessing the long-range coherence behaviour, the 1D polariton condensate in our experiments may roughly be characterized as a finite area but uniform 1D condensate due to its smooth steady-state profile.  It is now well appreciated that the CW-pumped polariton condensate may be viewed as confined to an 'effective trap' \cite{Roumpos2010, Ostrovskaya2012} within which excitations over the entire condensate may be observed. Thus, despite ballistic expansion, the spatially extended steady-state profile represents the system size, and we may qualitatively apply the expected 1D correlation functions to this experimental system.

%As shown in Fig. \ref{fig:Fig02}d), for a polariton wire in the condensate regime at a lattice temperature of 5 K, we can evidence the persistence of the ODLRO over a much larger distance in our microwire, which is indicated by the light blue area in Fig. \ref{fig:Fig02}d). As we show later in the manuscript, this distance can be in excess of 45\,$\upmu$m, and it features a stable and almost constant plateau in the correlation function which only starts to drop to the noise level towards the outer edge of the wire. This is in good qualitative agreement with theoretical predictions \cite{SavenkoPRB2011} and earlier reports on one-dimensional polariton condensates.

\begin{figure}[tbp]
\centering
\includegraphics[width=\linewidth]{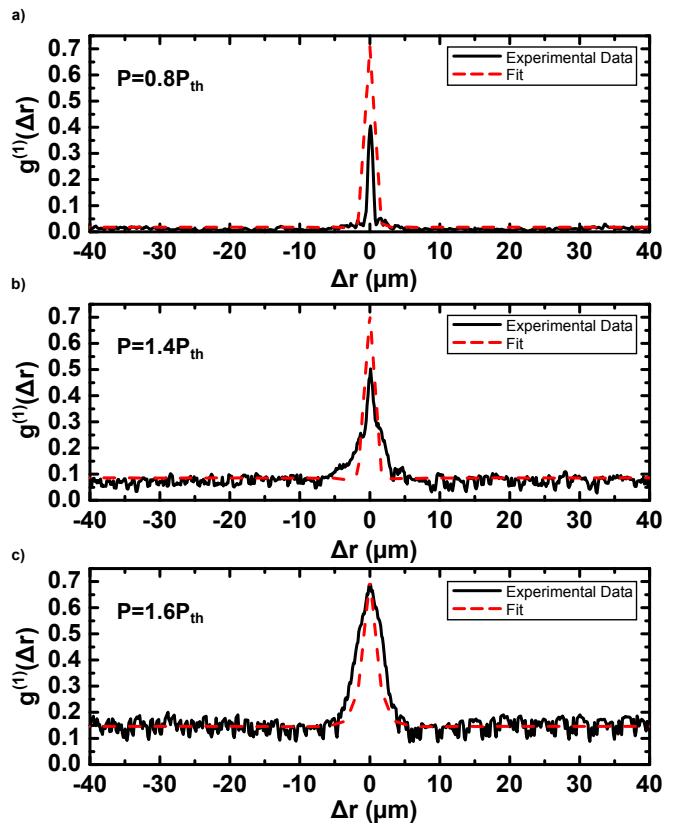}
\caption{$g^{\left(1\right)}\left(\Delta r\right)$-for different excitation powers a) below condensation threshold $P=0.8 P_{th}$, b) slightly above threshold $P=1.4P_{th}$ and c) $P=1.6P_{th}$. With higher excitation-powers the central peak width around $\Delta r = 0$ increases and the value of the $g^{\left(1\right)}\left(\Delta r\to \infty \right)$-function reaches a plateau of about 0.15 at $P=1.6P_{th}$. The red dashed lines in a)-c) are the results of our theoretic modeling.}
\label{fig:Fig03}
\end{figure}

\begin{figure}[tbp]
\centering
\includegraphics[width=\linewidth]{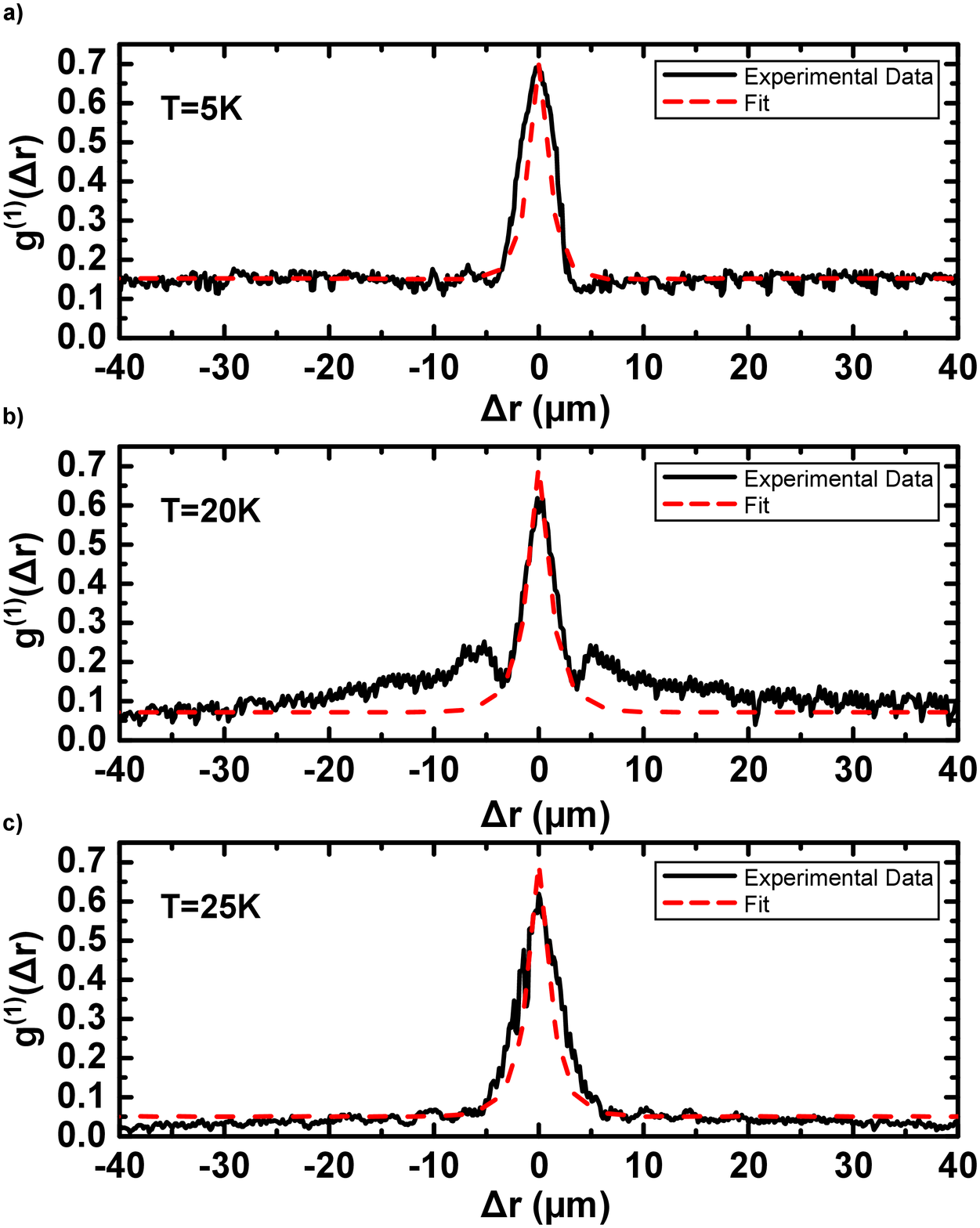}
\caption{$g^{\left(1\right)}\left(\Delta r\right)$-function for different sample temperatures a) $T=5$\,K, b) $T=20$\,K and c) $T=25$\,K at an excitation power of $P\approx 1.8 P_{th}$. The red dashed lines are again the results of our model. With increasing temperatures the plateau of $g^{\left(1\right)}\left(\Delta r\to \infty \right)$ disappears (c) $T=25$\,K.}
\label{fig:Fig04}
\end{figure}

{\it Theory. ---} Theoretically, the first order spatial coherence can be determined as
\begin{equation}
g^{\left(1\right)}(\textbf{r}',\textbf{r}'')=\frac{\chi(\textbf{r}',\textbf{r}'')}
{\sqrt{\chi(\textbf{r}',\textbf{r}')\chi(\textbf{r}'',\textbf{r}'')}},
\label{CoherenceDef}
\end{equation}
where $\chi(\textbf{r}',\textbf{r}'')$ is the single-particle density matrix of the system in real space:
\begin{equation}
\chi(\textbf{r}',\textbf{r}'')=\langle\hat{\Psi}^\dagger(\textbf{r}')\hat{\Psi}(\textbf{r}'')\rangle
\equiv\text{Tr}[\rho\hat{\Psi}^\dagger(\textbf{r}')\hat{\Psi}(\textbf{r}'')].
\end{equation}
Here $\hat{\Psi}^\dagger(\textbf{r}),\hat{\Psi}(\textbf{r})$ are exciton-polariton field operators, $\rho$ is a full density matrix of the closed system (polariton system plus the environment). 

In the case of a spatially homogeneous system, the first-order coherence is only dependent on the relative distance: $g^{\left(1\right)}\left(\textbf{r}',\textbf{r}''\right)=g^{\left(1\right)}\left(r\right)$, where $r=|\textbf{r}'-\textbf{r}''|$. In the ODLRO phase the correlations do not decay even at very large distances, and thus $\text{lim}_{r\rightarrow\infty}g^{(1)}(r)\neq0$ \cite{LeggettBook}.
In order to calculate $g^{\left(1\right)}\left(r\right)$ at different temperatures and pump intensities, we will employ a closed system of dynamic equations for the elements of the single particle density matrix in the reciprocal space based on the Lindblad master equation techniques and accounting for the processes of polariton-polariton and polariton-phonon interactions, pumping, and finite lifetime. The corresponding formalism was developed by us, and is described in greater detail elsewhere \cite{SavenkoPRB2011,MagnussonPRB2011}. In those previous works, we treated pulsed excitation of the condensate simply by introducing appropriate initial conditions \cite{SavenkoPRB2011}, or we considered CW resonant injection \cite{MagnussonPRB2011}. To accurately account for the effects of incoherent pumping, we had to extend our model by a stochastic approach introducing the random phase in the pumping term and performing statistical averaging. This modification represents a nontrivial and vitally important extension of our previous works to realistically model the experimental data. Upon finding the single-particle density matrix in the reciprocal space, the real space behavior can be found by the Fourier transform. The theoretical formalism is summarized in the Supplementary material \cite{Supplement} and full details can be found in Refs. \cite{SavenkoPRB2011,MagnussonPRB2011,SavenkoJETP2012}.

In the calculations, we used parameters (effective mass of polaritons, the Rabi splitting energy and detuning; the temperatures for each set of data; dimensions of the quantum wire) taken from the experimental data. The matrix element of polariton-polariton interaction was estimated using the expression: $U\approx 3E_ba_B^2/S$, where $E_b$ is the exciton binding energy, $a_B$ is its Bohr radius and $S$ is the area of the wire, which gave $U\approx 4$\,neV. The polariton-polariton scattering rate was taken independent of the wavevector, for simplicity. The polariton-phonon scattering rate is calculated using Eq. (7) in \cite{Supplement} and the parameters there. The maximum value of the scattering rate reads $W\approx 38$\,neV. Pumping powers were chosen in accordance with the experimental data.

\emph{Results and discussion.---} First, we investigate the behavior of coherence with the increase of the condensate density controlled by the intensity of pump, $P$. Below the threshold of condensation, the coherence function, $g^{\left(1\right)}\left(r\right)$, represents a resolution-limited sharp peak and a fast drop down to the noise level, which is shown in Fig. \ref{fig:Fig03}a). After reaching the polariton lasing threshold $P_{th}$ (Fig. \ref{fig:Fig03}b), this central peak starts to broaden indicating the increase of the coherence (Fig. \ref{fig:Fig03}c). In thermal equilibrium, the width of the peak is related to the thermal de Broglie wavelength $\lambda_{dB}$ of the Bose gas. Despite our system is clearly out of equilibrium, we follow this analogy and extract an effective de Broglie wavelength of $\lambda_{dB}\approx8.5$\,$\upmu$m well above the condensation threshold, which is in good agreement with earlier reports \cite{Roumpos-PNAS12}. More remarkable, the spatial correlation function acquires a nearly constant plateau in the regime of polariton condensation, which persists over the full range of the microwire significantly above the noise level. 

For moderate pump powers (below two times the threshold power) we observe a consistent increase of the spatial coherence degree with the increase of polariton occupancy. This is exemplarily demonstrated in Fig. \ref{fig:Fig03}c) for a pump power of $1.6 P_{th}$. We can quantitatively reproduce this behavior with our theory, as shown in Fig. \ref{fig:Fig03}a-c):
The spatial coherence is monotonously decaying over the distance $r$ and reaches some non-zero value $g_{\infty}^{\left(1\right)}$ for   $r\rightarrow\infty$. This value is determined by the percentage of the coherent fraction in the system and it increases with the pump power, $P$  (in case of absent strong localization effects). This behavior is also in good agreement with the theory.

Fig. \ref{fig:Fig04} serves as a manifestation of the thermal effects influence on the long range order. The figure depicts evolution of $g^{\left(1\right)}\left(r,-r\right)$ as a function of temperature at a constant pump power of 1.8\,$P_{th}$. The sample temperature is increased from 5 to 20 K.

 We observe a significant decrease of $g^{\left(1\right)}\left(r,-r\right)$ with increase of the temperature, which strongly indicates the detrimental influence of acoustic phonons on the spatial coherence properties of the condensate. While for the highest temperature the correlation function drops to the noise level within the size of the pump spot in almost perfect agreement with theory (Fig.~\ref{fig:Fig04}c), the occurrence of satellite peaks (Fig.~\ref{fig:Fig04}b, Fig.~\ref{fig:Fig03}b) indicate some persisting influence of sample disorder \cite{Trichet2013} or a boundary mode reflection.
 At a temperature of $T=25$\,K [Fig. \ref{fig:Fig04}c)] the value of $g^{\left(1\right)}\left(r,-r\right)$ quenches to zero at distances $r>20$\,$\upmu$m and transition from ODLRO to thermal phase occurs.

%--------------------------------------------- %---------------------------------------------
%---------------------------------------------

\emph{Conclusion.---} To summarize, we have investigated theoretically and experimentally spatial coherence properties of a one-dimensional exciton-polariton microwire under different non-resonant pump intensities and temperatures. We have compared the experimental data with calculations, utilizing a stochastic master equation approach. It has been demonstrated that the $g^{\left(1\right)}\left(r\right)$-function has a plateau region which is determined by the intensity of non-resonant pumping and the effective temperature of the crystal lattice. Our results indicate that the method of stochastic phase in the framework of the master equation approach allows to simulate incoherent pumping of the system.

The authors would like to thank the State of Bavaria for
financial support. I.A.S acknowledges support of FP7 IRSES project POLAPHEN. I.G.S. was partially supported by the Academy of Finland through its COMP (project no. 251748) and LTQ (project no. 250280) Centre of Excellence grants and the Government of Russian Federation, Grant 074-U01.

%------------------------------------------------------------------------------------------------------------
%------------------------------------------------------------------------------------------------------------
%------------------------------------------------------------------------------------------------------------
%------------------------------------------------------------------------------------------------------------
%------------------------------------------------------------------------------------------------------------

\end{document}

\title{SUPPLEMENTARY MATERIALS --
Spatial Coherence Properties of One-Dimensional Exciton-Polariton-Condensates}

\author{J. Fischer}
\affiliation{Technische Physik, Wilhelm-Conrad-R\"ontgen-Research Center for Complex Material Systems, Universit\"at W\"urzburg, Am Hubland, D-97074 W\"urzburg, Germany}

\author{I. G. Savenko}
\affiliation{QCD Labs, COMP Centre of Excellence, Department of Applied Physics and Olli V. Lounasmaa Laboratory, Aalto University, P.O. Box 13500, FI-00076 Aalto, Finland}
\affiliation{National Research University of Information Technologies, Mechanics and Optics (ITMO), St. Petersburg 197101, Russia}

\author{M. D. Fraser}
\affiliation{Quantum Optics Research Group, RIKEN Center for Emergent Matter Science, 2-1 Hirosawa, Wako-shi, Saitama 351-0198, Japan}

\author{S. Holzinger}
\affiliation{Technische Physik, Wilhelm-Conrad-R\"ontgen-Research Center for Complex Material Systems, Universit\"at W\"urzburg, Am Hubland, D-97074 W\"urzburg, Germany}

\author{S. Brodbeck}
\affiliation{Technische Physik, Wilhelm-Conrad-R\"ontgen-Research Center for Complex Material Systems, Universit\"at W\"urzburg, Am Hubland, D-97074 W\"urzburg, Germany}

\author{M. Kamp}
\affiliation{Technische Physik, Wilhelm-Conrad-R\"ontgen-Research Center for Complex Material Systems, Universit\"at W\"urzburg, Am Hubland, D-97074 W\"urzburg, Germany}

\author{I. A. Shelykh}
\affiliation{Science Institute, University of Iceland, Dunhagi 3, IS-107, Reykjavik, Iceland}
\affiliation{Division of Physics and Applied Physics, Nanyang Technological University 637371, Singapore}

\author{C. Schneider}
\affiliation{Technische Physik, Wilhelm-Conrad-R\"ontgen-Research Center for Complex Material Systems, Universit\"at W\"urzburg, Am Hubland, D-97074 W\"urzburg, Germany}

\author{S. H\"ofling}
\affiliation{Technische Physik, Wilhelm-Conrad-R\"ontgen-Research Center for Complex Material Systems, Universit\"at W\"urzburg, Am Hubland, D-97074 W\"urzburg, Germany}
\affiliation{SUPA, School of Physics and Astronomy, University of St Andrews, St Andrews KY16 9SS, United Kingdom}

\date{\today}
\maketitle

%\pacs{42.55.Sa, 78.67.De, 78.20.Ls, 78.55.Cr}

\section*{Sample details}

The polariton channels under investigation with a length of 200 µm (width 5 µm) were etched into a high-Q AlGaAs based $\lambda/2$ microcavity with twelve GaAs quantum wells (QWs). The QWs are located in the optical antinodes, in a similar manner as described in \cite{Deng-Science02}. The Q-factor is experimentally determined to exceed 10000. In order to generate a quasi one-dimensional polariton channel, microwires with the length of 200\,$\upmu$m and width of 5\,$\upmu$m were created via electron beam lithography and etched deeply into the structure using electron-cyclotron-resonance reactive-ion-etching. Due to the optimized etching technique resulting in smooth and steep sidewalls with extremely small roughness, no detrimental influence on the cavity Q-factor was caused. A scanning electron microscope image of such a wire structure is shown in Fig. 1a).

\section*{Experimental details}

In Fig. 2a) of the main text we schematically show the experimental setup for the measurement of the $ g^{\left(1\right)}\left(r,-r\right) $-function. The signal P originating from the sample is divided in two parts by a $ 50:50 $ non-polarizing beamsplitter. The signal which enters first arm is reflected by a mirror (P) and it is directly imaged on the CCD. At the end of the second arm, which is mounted on a combination of a translation stage and a piezo-actuator, we installed is a prism instead of a mirror, which produces in realspace a flipped image (\reflectbox{P}) of the sample-emission pattern. On the CCD the two images are overlapped and one receives the interference-image, which is shown in Fig. 2b) of the main text. The overlapping realspace coordinates are now $ \left(r,y\right) $ and $ \left(-r,y\right) $. By changing the path difference $ x $ between the two interferometer arms with the help of the piezo-actuator the interference fringes are shifted on the CCD-image. The intensity of each pixel has a sinusoidal modulation dependent on the path difference $ x $. By recording images for different path delays $ x $ up to a maximum path difference of about $ x_{max} = 2\lambda_{con} $ ($ \lambda_{con} $: emission wavelength of the condensate), which is in our case about $ x_{max}\sim 1.6 $\,$ \upmu $m, we can extract the sinusoidal intensity modulation for every pixel of the CCD-image  (main text Fig. 2c)). We fit this data with the following sinus-function:

\begin{equation}
I\left(x\right) = I_{0} + A \sin\left(\frac{2\pi}{\lambda_{con}}\left(x-x_{0}\right)\right)\,\text{,}
\end{equation}
where $ x $ is the path difference between the two arms of the interferometer. From the fitted function we can directly extract the $ g^{\left(1\right)}\left(r,y,-r,y\right) $-value $ g^{\left(1\right)}\left(r,y,-r,y\right)= g^{\left(
1\right)}\left(\Delta r,y\right)=\frac{A}{I_{0}} $ and the relative phase $ \phi = \frac{2\pi x_{0}}{\lambda_{con}} $. By performing this analysis for every pixel of the image, we can extract the coherence map of Fig. 2d). For the comparison of the theory with the experiment we extract a cross-section of the coherence map at the center along the length of the polariton-wire ($ y = 0 $).

\begin{figure}[tbp]
\centering
\includegraphics[width=12cm]{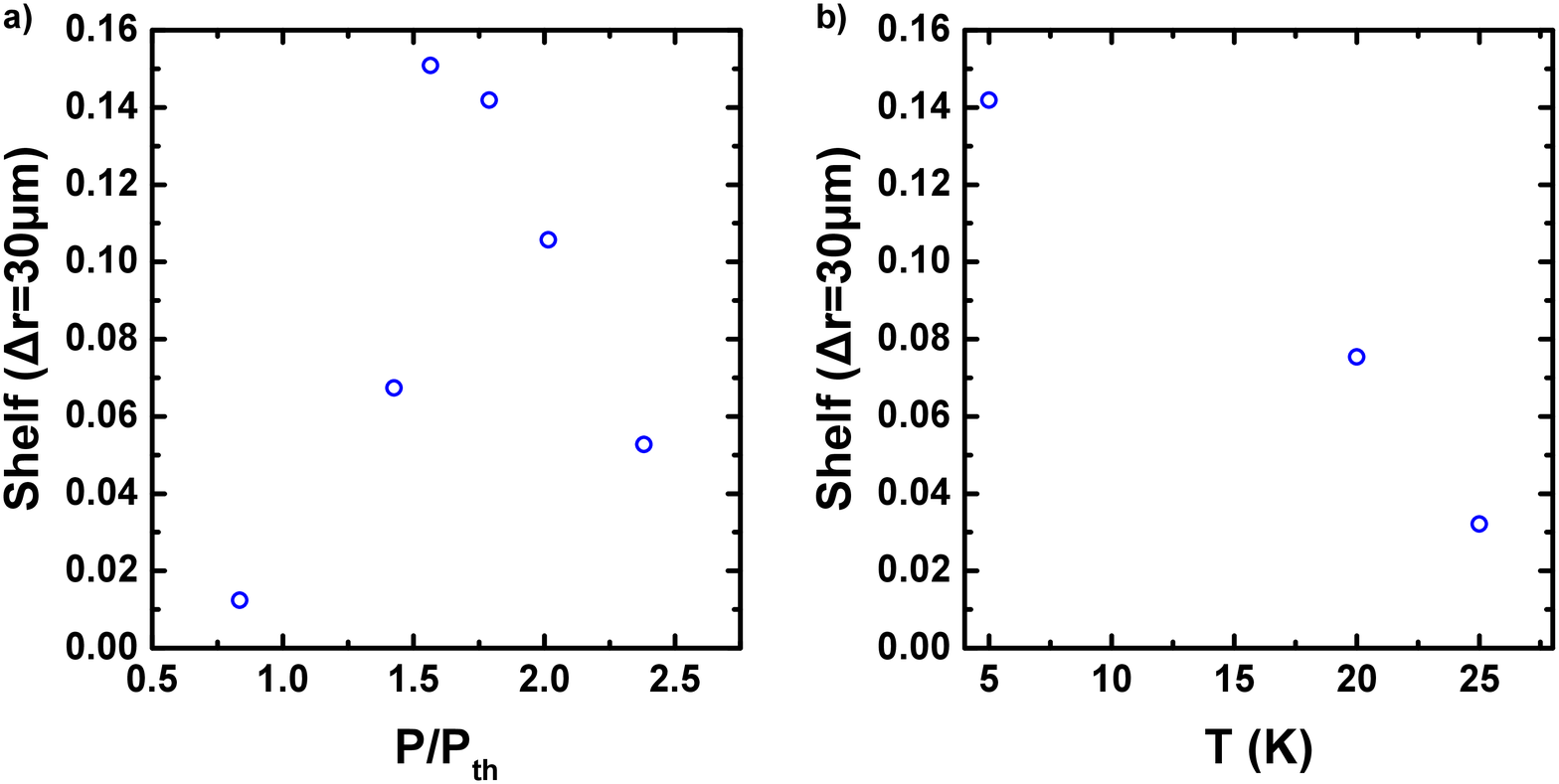}
\caption{a) and b) show the $g^{\left(1\right)}\left(\Delta r = 30\text{\,}\upmu\text{m}\right)$-value of the correlation function at a spatial separation of 30\,$\upmu$m away from the pump spot as a function of the pump power or sample temperature, respectively.}
\label{fig:SupplFig1}
\end{figure}

In Suppl. Fig. \ref{fig:SupplFig1}a) and b) we plot the $g^{\left(1\right)}\left(\Delta r = 30\text{\,}\upmu\text{m}\right)$-value extracted 30\,$\upmu$m away from the excitation spot as a function of the external parameters, pump power and sample temperature. Below threshold the height of the shelf is the noise level of the system (about 0.03-0.04), above threshold at $P>P_{th}$ the shelf height increases up to 0.15 (Suppl. Fig. \ref{fig:SupplFig1}a)). It is worth noting, that the drop of the $g^{\left(1\right)}\left(30\text{\,}\upmu\text{m}, −30\text{\,}\upmu\text{m}\right)$ value for larger pump powers is most likely caused by power-induced sample heating, which again confirms the detrimental impact of thermally activated phonons on the coherence properties of the condensate. The heating is also seen in Fig. 1e) of the main text, due to the intensity decrease at higher excitation powers. In Suppl. Fig. \ref{fig:SupplFig1}b) it is clearly seen, that the shelf decreases with temperature and at $T=25$\,K the value is again at the noise level of our setup.

\section*{Phase fluctuations in a 1D polariton condensate}

\noindent
In a 1D Bose gas \cite{petrov2000regimes}, in the weakly interacting regime, the expectation value for phase fluctuations can be represented as 
\begin{equation}
   \langle \delta\phi^{2}(x)\rangle \propto \frac{|x|}{l_{\phi}}
\end{equation}

\noindent
where 
\begin{equation}
    l_{\phi}=\frac{\hbar^{2}n_{s}}{mk_{B}T} 
\end{equation}

\noindent
is the 1D coherence length which characterises the expected length over which coherence decays as a result of phase fluctuations, i.e. for a quasi-condensate, one expects an exponential decay $g^{1}(x)\propto\exp(-|x|/l_{\phi})$.  With some realistic values in place for the sample parameters, such as sample temperature $T=4K$ and polariton mass $m=0.2\times10^{-4}m_{e}$, where $m_{e}$ is the vacuum electron mass, it is found that even for a small 1D density of $n_{s}=10\upmu m^{-1}$, the coherence length is found to be $l_{\phi}>100\upmu m$.  The typical system size $R$, i.e. the steady-state length over which the condensate density exceeds the critical density, is thus generally less than $l_{\phi}$.  The inequality $R<l_{\phi}$ implies the suppression of phase fluctuations for this parameter space location, and further, that the condensate should be closer to a true-BEC than a quasi-BEC state and exhibit plateauing $g^{1}(r)$ at large distances, precisely that found in our experiments. Enhancing the 1D density will further enhance the coherence length and the case for characterizing the 1D polariton BEC as a true-BEC.\\

\section*{Theoretical approach details}

We describe the system by the full 
density matrix factorizing it into polariton and phonon parts (Born approximation) $\rho=\rho_{ph} \otimes \rho_{pol}$, where the phonon part of the system is assumed to be thermalized and time- independent, $\rho_{ph}=\texttt{exp}\left\{-\beta {H}_{ph}\right\}$. 
The single-particle density matrix in real space is a Fourier transform of the density matrix in reciprocal space:
\begin{eqnarray}
\chi(\textbf{r}',\textbf{r}'')=\left(\frac{2\pi}{L}\right)^d\int
e^{-i(\textbf{k}'\textbf{r}'-\textbf{k}''\textbf{r}'')}\chi(\textbf{k}',\textbf{k}'')d\textbf{k}'d\textbf{k}'',
\label{Fourier}
\end{eqnarray}
where
\begin{equation}
\chi(\textbf{k}',\textbf{k}'')=\langle a^\dagger_{\textbf{k}'}a_{\textbf{k}''}\rangle\equiv\text{Tr}[\rho a^\dagger_{\textbf{k}'}a_{\textbf{k}''}],
\end{equation}
$d$ is the dimensionality of the system ($d=1$ for a 1D microwire), $L$ is the characteristic size of the system (length of the wire in our case) and $a^\dagger_{\textbf{k}},a_{\textbf{k}}$ are the creation and annihilation operators for the polaritons with momentum \textbf{k}.
The diagonal elements of the single-particle density matrix in reciprocal space give occupation numbers of the states with given \textbf{k}.

The dynamics of the polariton part of the density matrix is governed by the Hamiltonian of the system, which can be represented in the following form:
\begin{eqnarray}
\nonumber
H &=& \sum_{\textbf{k}} E_\textbf{k} a_{\textbf{k}}^\dagger a_\textbf{k}+ \sum_{\textbf k} \left(p_{\textbf k} a_{\textbf k}^\dagger + p^*_{\textbf k} a_{\textbf k}\right)\\
\nonumber
&+& \sum_{\textbf{k},\textbf{k}',\textbf{p}}\frac{U_{\textbf{k},\textbf{k}'}(\textbf{p})}{2}a_{\textbf{k}}^\dagger a_{\textbf{k}'}^\dagger a_{\textbf{k}+\textbf{p}}a_{\textbf{k}'-\textbf{p}} \\
\nonumber
&+& \sum_{\textbf{k}, \textbf{q}}D_{\textbf{k}}(\textbf{q})a_{\textbf{k}+\textbf{q}}^\dagger a_\textbf{k}(b_\textbf{q}+b_{-\textbf{q}}^\dagger),
\label{EqHamiltonian}
\end{eqnarray}
where the first term describes free-polariton kinetic energy, the second term corresponds to the pumping, the third term accounts for the polariton-polariton scattering, the fourth term corresponds to the incoherent processes of interaction between polaritons and acoustic phonons.

Here $E_\textbf{k}$ is the free-polariton dispersion, parameters $U(\textbf{p})$ and $D(\textbf{q})$ correspond to the polariton-polariton and polariton-phonon scattering rates. They can be obtained for corresponding parameters for 2D excitons (see \cite{YamamotoTassone_1999,Piermarocchi}) by multiplication by excitonic fractions of polariton states (Hopfield coefficients) \cite{SolnyshkovDisp}:
\begin{equation}
U_{\textbf{k},\textbf{k}'}(\textbf{p})\approx 6E_b\frac{a_B^2}{S}X_{\textbf{k}}X_{\textbf{k}'}X_{\textbf{k+p}}X_{\textbf{k}' -\textbf{p}},
\end{equation}
\begin{equation}
D_{\textbf{k}}(\textbf{q})\approx i\sqrt{\frac{\hbar\left(|\textbf{q}|^2+q_z^2\right)^{\frac{1}{2}}}{2\rho_{m} Vc_s}}\left(d_eI_e-d_hI_h\right)X_\textbf{k}X_{\textbf{k}+\textbf{q}},
\end{equation}
where in the first formula $E_b$ and $a_B$ the exciton binding energy and the Bohr radius, $X_\textbf{p}$ are the excitonic Hopfield coefficients (excitonic fractions in polaritons for the given wavevector $\textbf{p}$), $q_z=2\pi/w$ is perpendicular to the plane of the QW component of the wavevector and $w$ is the QW width. In the second formula, $\rho_{m}$ is the density of the material of the QW, $V=Ldw$ is the quantization volume, $c_s$ is the sound velocity in the corresponding material. Parameters $d_e$ and $d_g$ are the deformation potentials for electrons and holes. Symbol $I$ corresponds to the superposition integrals of the excitonic and phononic wavefunctions in the QW.
%
%\begin{equation}
%\nonumber
%I_{e(h)}\approx \left[1+\left(\frac{m_{e(h)}}{2(m_e+m_h)}|\textbf{q}|a_B\right)^2\right]^{-3/2}.
%\end{equation}
%

The coefficients $p_{\textbf{k}}$  are the pump intensities. The physical meaning of the coherent pumping is coupling of the system to an electromagnetic field with a well-defined phase, provided, for instance, by a laser beam focused on the quantum wire. However, in current work we apply a stochastic phase modulation, thus making the pumping scheme quasi-non-resonant. The coefficients $p_\textbf k$ are Fourier transforms of the pumping amplitudes in a real space $p(\textbf x,t)$ which in the case of $cw$ pump can be cast as:
\begin{equation}
p(\textbf x,t)= P(\textbf x)e^{i\textbf k_p \textbf x} e^{-i\omega_p t},
\end{equation}
where $P(\textbf x)$ is the pumping spot profile in real space. Since we consider homogenous pumping, $P(\textbf x)=P_0=const(\textbf x)$; $\textbf k_p$ is an in-plane pumping vector (the angle of inclination of the laser beam w.r.t.~the sample) and $\omega_p$ is the pumping frequency, which is a stochastic variable in our case ($\tilde\omega_p$). The wavevector is connected with the pumping frequence by the dispersion relation. Therefore, $\textbf k_p=\textbf k_p(\tilde\omega_p) = \tilde k_p$ in our 1D case. Thus, the pumpimg term takes the form:
\begin{equation}
p(x,t)= P_0e^{i\tilde k_p x} e^{-i\tilde\omega_p t}.
\end{equation}

Due to the fact that initially coherent pumping terms have stochastic nature in our approach (having random pumping energy at each iteration), these terms effectively play a role of incoherent pumping. In this sense, the theoretically used pumping scheme is very close to the experimental setup, where coherent but frequency-detuned pumping was used. Moreover, the results of the experiment and theoretical data are in  good agreement. Therefore, we consider such theoretical trick justified.

In the Hamiltonian~\eqref{EqHamiltonian} we can distinguish between the part $H_{h}$ describing the Hamiltonian evolution of the system and including the terms of the kinetic energy, pump and polariton-polariton interactions, and the part $H_{th}$ describing the processes of thermalization coming from polariton-phonon interactions,  $H=H_{h}+H_{th}$. The dynamics governed by these two parts should be treated independently, and total equation for the density matrix can be represented in the form of the master equation accounting for both types of the processes:
\begin{equation}
\partial_t\rho=\left(\partial_t\rho\right)^{(h)}+\left(\partial_t\rho\right)^{(th)},
\end{equation}
where the coherent part of the evolution is governed by the Liouville-von Neumann equation:
\begin{equation}
i\hbar\left(\partial_t\rho\right)^ {(h)} =\left[H_{h};\rho\right],
\label{liouville}
\end{equation}
and incoherent part can be represented in the Lindblad form \cite{Carmichael}:
\begin{eqnarray}
\nonumber
\left(\partial_t\rho\right)^{(th)}=-\frac{1}{\hbar^2}\int_{-\infty}^t \left[H_{th}(t);
 \left[H_{th}(t');\rho(t)\right]\right]dt' \\
%\label{Liouville_int}\\
\nonumber
=\delta_{\Delta E}\left[2\left(H^+\rho H^-+H^-\rho\right.H^+\right) \\
%\nonumber
\left.-\left(H^+H^-+H^-H^+\right)\rho-\rho\left(H^+H^-+H^-H^+\right)\right],
\label{Lindblad}
\end{eqnarray}
where the coefficient $\delta_{\Delta E}$ denotes the energy conservation, the terms $H^+$ and $H^-$ correspond to the processes of creation or destruction of the external reservoir particle, $H_{th}=H^++H^-$,$H^+=\sum_{\textbf{k}, \textbf{q}}D_{\textbf{k}}(\textbf{q})a_{\textbf{k}+\textbf{q}}^\dagger a_\textbf{k}b_{-\textbf{q}}^\dagger$.

Further, we obtain the full system of equations for the dynamics of the polariton system by combining the equations above and finding the expression for the single-particle density matrix:
\begin{equation}
\partial_t \chi(\textbf{k},\textbf{k}')=Tr\left\{\left(\partial_t\rho\right) a_{\textbf{k}}^\dagger
a_{\textbf{k}'}\right\}\equiv\partial_t\langle a_{\textbf{k}}^\dagger
a_{\textbf{k}'}\rangle,
\end{equation}
where the trace is performed by all the degrees of freedom of the system. Thus, the dynamics equation for the diagonal and off-diagonal elements of the single-particle density matrix reads:
\begin{eqnarray}
\nonumber
\left\{\partial_t\chi(\textbf{k},\textbf{k}')\right\}
=\frac{i}{\hbar}(E_\textbf{k}-E_{\textbf{k}'})\chi(\textbf{k},\textbf{k}')\\
\nonumber
+\frac{i}{\hbar}\left( p_{\textbf k_1}^* \langle a_{\textbf
k_2}\rangle-p_{\textbf k_2} \langle a_{\textbf k_1}
\rangle^*\right)
-\left(\frac{1}{2\tau_{\textbf{k}}}+\frac{1}{2\tau_{\textbf{k}'}}\right)\chi(\textbf{k},\textbf{k}')\\
\nonumber
+i\frac{U}{\hbar}\sum_{\textbf{k}_1,\textbf{p}}\chi(\textbf{k}_1,\textbf{k}_1-\textbf{p})\left[\chi(\textbf{k}-\textbf{p},\textbf{k}')-\chi(\textbf{k},\textbf{k}'+\textbf{p})\right]\\
\nonumber
\chi(\textbf{k},\textbf{k}')\left\{\sum_{\textbf{q}',E_\textbf{k}<E_{\textbf{k+q}'}}W(\textbf{q}')\left[\chi(\textbf{k+q}',\textbf{k+q}')-n_{\textbf{q}'}^{ph}\right]+
\right. \\
\nonumber
+\sum_{\textbf{q}',E_\textbf{k}>E_{\textbf{k+q}'}}W(\textbf{q}')\left[-\chi(\textbf{k+q}',\textbf{k+q}')-n_{\textbf{q}'}^{ph}-1\right]+\\
\nonumber
+\sum_{\textbf{q}',E_{\textbf{k}'}<E_{\textbf{k}'+\textbf{q}'}}W(\textbf{q}')\left[\chi(\textbf{k}'+\textbf{q}',\textbf{k}'+\textbf{q}')-n_{\textbf{q}'}^{ph}\right]
+\\
%\nonumber
+\left.\sum_{\textbf{q}',E_{\textbf{k}'}>E_{\textbf{k}'+\textbf{q}'}}W(\textbf{q}')\left[-\chi(\textbf{k}'+\textbf{q}',\textbf{k}'+\textbf{q}')-n_{\textbf{q}'}^{ph}-1\right]\right\},
\label{EqnFullOffdiag}
\end{eqnarray}
where the first line in the formula accounts for the free-particles kinetic energy, the second one is responsible for the pump and finite lifetime, the third line describes polariton-polariton scattering, and the rest of the equation refers to the polaritons interaction with the reservoir of acoustic phonons. Here $n_\textbf q^{ph}$ is the number of phonons with momentum $\textbf q$, given by the Bose distribution, and $W(\textbf{q}')$ are the transition rates of phonon-assisted processes.

It can be seen from the formula above that in order to account for the coherent pump it is required to know the dynamics of the order parameters $\langle a_{\textbf{k}}\rangle$. Therefore, we make our system of equations complete by adding $N$ additional equations:
\begin{eqnarray}
\label{eq:pumping1}
&&\partial_t \langle a_\textbf k \rangle =\\
\nonumber
&-&\frac{i}{\hbar} E_\textbf k \langle a_\textbf k \rangle
-\frac{i}{\hbar} p_\textbf k\\
\nonumber
 &-&\frac{i}{\hbar} U \sum_{\textbf {q},\textbf p} \chi(\textbf {k+q},\textbf
{k+q}-\textbf p) \langle a_{\textbf k+\textbf p} \rangle\\
\nonumber
 &+&\left\{\sum_{\textbf q,E_\textbf k<E_{\textbf k+\textbf q}} W(\textbf
q) (\chi(\textbf k+\textbf q,\textbf k+\textbf q) - n_\textbf q^{ph})
\right.\\ \nonumber
&+&\left. \sum_{\textbf q,E_\textbf k>E_{\textbf k+\textbf q}} W(\textbf
q) (-\chi(\textbf k+\textbf q,\textbf k+\textbf q) - n_\textbf
q^{ph}-1)\right\}\langle a_\textbf k \rangle.
\end{eqnarray}

Coupled equations (14) and (15) represent a nonlinear convergent system, first-order in time. Therefore, a standard Runge-Kutta numerical iterative scheme was employed in order to find the consistent solutions. The initial condition corresponded to zero polariton occupation in the sample and absent polariton-phonon coupling. The correlators and mean values of the creation-annihilation operators were also set to zero.